\documentstyle[prb,aps,epsf,multicol]{revtex}

\begin{document}

\title{Elastic Constants of Quantum Solids by Path Integral Simulations}

\author{Philipp Sch\"offel and Martin H. M\"user}

\address{Institut f\"ur Physik, WA 331; Johannes Guntenberg-Universit\"at\\              55099 Mainz; Germany}

\date{\today}
\maketitle

\begin{abstract}
Two methods are proposed to evaluate
the  second-order elastic constants of quantum mechanically treated solids.
One method is based on path-integral simulations
in the $NVT$ ensemble using an estimator for elastic constants 
$C_{ij}$. 
The other method is based on simulations in the
$NpT$ ensemble exploiting the relationship between strain fluctuations
and elastic constants.
The strengths and weaknesses of the methods are discussed thoroughly.
We show how one can reduce statistical and systematic errors associated
with so-called primitive estimators.
The methods are then applied to solid argon at atmospheric pressures
and solid helium 3 (hcp, fcc, and bcc) under varying pressures. 
Good agreement with available
experimental data on elastic constants
is found for $^3$He. Predictions are made for the thermal expectation
value of the kinetic energy of solid $^3$He.
\end{abstract}
\vspace*{2mm}

\hspace*{16mm}{PACS: 67.80.-s, 62.20.Dc, 02.70.Ns, 05.30.-d}

\begin{multicols}{2}

\section{Introduction}
The quantum nature of atomic motion alters thermal and mechanical properties
of condensed matter at low temperatures. The most striking feature is the 
well-known quantum mechanical freezing of solids below the Debye 
temperature: Specific heat and thermal expansion vanish as the 
temperature approaches absolute zero~\cite{ashcroft76}.
This behavior is different from what
we would expect from classical statistical mechanics. A ``classical solid''
typically has a positive expansion coefficient at low temperatures
and a specific heat close to $3 k_B$ per atom.

Path integral Monte Carlo (PIMC)~\cite{berne86,schmidt95,chakravarty97,marx99}
and path integral molecular dynamics (PIMD)~\cite{tuckerman93,tuckerman98}
are convenient numerical methods to predict the proper low-temperature
properties of condensed matter, provided that model potentials are available
that describe the interatomic interactions sufficiently well.
So far, most PIMC and PIMD studies have been restricted to the
calculation of structural and thermal properties of quantum solids or to
the calculation of equations of state of condensed rare gases.
The computation of the tensor of the elastic constants, which is
an important material property in many engineering applications, however,
has not been put forward yet in the framework of path-integral simulations.

The proper evaluation of the elastic constants is far more complicated than
textbooks on solid state physics make us believe, because it is not
sufficient to calculate second derivates of interaction potentials at
equilibrium (or thermal) positions and construct from this the various
$C_{ij}$.
As shown by Squire, Holt, and Hoover,~\cite{squire69}
thermal fluctuations of a generalized
strain tensor can alter elastic constants significantly.
Of course, quantum fluctuations can be expected to result in a similar effect.
These fluctuations vanish at small temperatures for classical systems, but
they do not necessarily vanish for quantum mechanical systems.
Another difficulty is that quantum solids do not
adopt a configuration at $T = 0$~K where the potential energy
surface is minimized. Instead, the atoms also probe non-harmonic
parts of the potentials due to quantum fluctuations, typically
leading to lattice constants that are slightly increased
with respect to the equivalent classical system.

In this paper, we want to propose two methods which enable us to
determine accurately elastic constants of solids such that quantum effects
of the atomic motion are fully included in the treatment. One possibility
is to use the definition of the elastic constants as the second derivative
of the free energy with respect to strain tensor elements and evaluate the
final expression in the $NVT$ ensemble. This procedure
is similar to the method proposed by Squire, Holt, and Hoover,~\cite{squire69}
with the difference
that our partition function is quantum mechanical. Alternatively, one can
perform simulations in the $NpT$ ensemble and relate the fluctuations of the
strain tensor to the elastic constants as has been originally done by 
Parrinello and Rahman for classical systems.~\cite{parrinello82}

Both methods are prone to produce large statistical error bars.
In classical simulations, the computation of elastic constants using
the $NpT$ ensembles is known to be much
less efficient than in the $NVT$ ensemble~\cite{sprik84}.
However, the estimator derived for the quantum simulations in the $NVT$
ensemble can be expected to become unreliable when the Trotter number $P$
in the path-integral simulation becomes large, just like the so-called
primitive estimator for the kinetic energy~\cite{herman82}.
It is thus necessary to discuss
the prospective statistical errors in detail.

It is important to note that presently only isothermal quantum mechanical
elastic constants can be calculated. Adiabatic elastic constants can
only be obtained if simulations are carried out in the $NVE$ ensemble or
the $NpH$ ensemble.~\cite{parrinello82}
Constraining the energy $E$ or the enthalpy $H$
to constant values in the simulation does not necessarily correspond to
conserved $E$ or $H$ of the real quantum system.

The remainder of this paper is organized as follows:
In Sec.~\ref{sec:theory}, the equations are derived that allow the
correct determination of all $C_{i j}$ in the $NVT$ ensemble
in terms of a path integral formulation. In this context, we will
give an in depth discussion of the statistical errors that will arise
as a consequence of the corrections to  $C_{i j}$
associated with the quantum mechanical kinetic energy. It will be
shown, how, simple corrections to primitive estimators can reduce
their statistical incertainties considerably.
We will also briefly
review the method to determine $C_{i j}$ in the $NpT$ ensemble
as well as the simulation algorithm that is
used for the applications.
In Sec.~\ref{sec:applications}, the methods will be applied to
the calculation of elastic constants of
solid Argon as well as of fcc, bcc, and, hcp $^3$He.
In the case of $^3$He, the corrections to primitive estimators
will also be used to make predictions for thermal expectation
values of the kinetic energy.
Conclusions are drawn in Sec.~\ref{sec:conclusions}.

\section{Theory}
\label{sec:theory}

\subsection{Derivation of Elastic Constants}
\label{sec:deriv}
As pointed out correctly for classical systems~\cite{squire69},
isothermal elastic constants~\cite{comment}
$C_{ij}$ are
insufficiently described if they are evaluated solely on the basis of
averageing the so-called Born contribution $C_{ij}^{\rm (Born)}$
\begin{equation}
C_{ij}^{\rm (Born)} = {1\over V} \left\langle
\partial^2 V_{\rm pot} \over \partial \epsilon_i \partial \epsilon_j
\right\rangle,
\label{eq:born}
\end{equation}
which is the thermal average
of the second derivative of the potential
energy $V_{\rm pot}$ with respect to strain tensor elements $\epsilon_i$
and $\epsilon_j$ in the $NVT$ ensemble. The full elastic constants
are obtained if $\langle V_{\rm pot}\rangle$ is replaced with the free energy
$F(N,V,T) = -k_B T \ln Z(N,V,T)$,
e.g.,
\begin{equation}
C_{ij} =  -{k_B T\over V} 
{\partial^2 ln Z(N,V,T) \over  \partial \epsilon_i \partial \epsilon_j}
\label{eq:ela_def}
\end{equation}
with $Z(N,V,T)$ the isothermal partition function.
This proper definition results in
corrections to $C_{ij}^{\rm (Born)}$.
The leading correction terms $C_{ij}^{\rm (fluc)}$
are fluctuations of the (generalized) instantaneous strains.
These terms have the form
\begin{equation}
C_{ij}^{\rm (fluc)} = {\beta \over V} \left(
\left\langle{\partial V_{\rm pot} \over \partial \epsilon_i}\right\rangle
\left\langle{\partial V_{\rm pot} \over \partial \epsilon_j}\right\rangle
\, - \,
\left\langle 
{\partial V_{\rm pot} \over \partial \epsilon_i}
{\partial V_{\rm pot} \over \partial \epsilon_j}
\right\rangle 
\right),
\label{eq:correct}
\end{equation}
where again the expectation values are evaluated in the $NVT$ ensemble.
As far as classical elastic constants are concerned, the only missing
contributions $C^{\rm (kin)}_{ij}$ to the correct $C_{ij}$ 
stem from the ideal gas part of the partition function. These kinetic
corrections are given by
\begin{equation}
C^{\rm (kin)}_{ij} = -{N k_B T \over V} 
{\partial^2 \ln V\over \partial \epsilon_i \partial \epsilon_j}
\label{eq:ideal}
\end{equation}
or in tensor notation for cubic symmetry
\begin{equation}
C^{\rm (kin)}_{\alpha\beta\gamma\delta} =
{N k_B T \over V} \delta_{\alpha\delta} \delta_{\beta\gamma}.
\label{eq:ideal_tensor}
\end{equation}
Note that one has to properly symmetrize when using the
Voigt notation, e.g., 
$C^{\rm (Voigt)}_{12} = (C_{1212} + C_{1221} + C_{2112} + C_{2121})/4$.
The originally suggested values~\cite{squire69} for 
$C_{11}^{\rm (kin)}$, $C_{44}^{\rm (kin)}$ and
their symmetry related $C_{ij}^{\rm (kin)}$
are in error by a factor of two. The correct values are
$C_{11}^{\rm (kin)} = Nk_BT/V$, $C_{44}^{\rm (kin)} = Nk_BT/2V$, and
$C_{12}^{\rm (kin)} = 0$.
While this kinetic contribution can usually be neglected in classical
simulations, knowing the correct form is crucial for quantum systems as
can be seen further below. Thus, in classical solids at non-zero temperaures,
the elastic constants can be estimated as 
$C_{ij} = C_{ij}^{\rm (Born)} 
+ C_{ij}^{\rm (fluc)} + C_{ij}^{\rm (kin)}$.

The quantum statistical formulation of the partition function of
$N$ identical particles with mass $m$ in terms
of a path-integral formulation~\cite{feynman65} is given by
\begin{eqnarray}
Z(N,V,T) & = & \lim_{P \to \infty} \lambda^{-3NP}(\beta/P)
\nonumber \\ 
& & \times
\int d^3 R_{11} \cdots \int d^3 R_{NP} 
\exp\left(-{\beta\over P} V_{\rm eff} \right)
\label{eq:part_func}
\end{eqnarray}
with $\lambda(\beta/P)$ the thermal de Broglie wavelength at temperature
$\beta/P$ and the effective potential $V_{\rm eff}$
\begin{equation}
V_{\rm eff} = \sum_{n=1}^N \sum_{t = 1}^P \left\{
{m \over 2} \left(
{\vec{R}_{i\,t} - \vec{R}_{i\,t+1} \over \beta\hbar/P} 
\right)^2 \, + \,
\sum_{n'>n} V_{\rm pot}\left( {R}_{nn'\,t} 
\right) 
\right\} .
\label{eq:pot_eff}
\end{equation}
Here, the coordinate of particle $n$ at ``Trotter time'' $t$ is
denoted as $\vec{R}_{nt}$ and
${R}_{nn'\,t}$ denotes the distance between particle $n$ and $n'$ at
``Trotter time'' $t$.
If exchange effects are neglected,
one quantum point particle is represented as a closed classical ring 
polymer with the boundary condition $\vec{R}_{n\,t} = \vec{R}_{n\,t+P}$.
The temperature at which the simulation of the representation of the
quantum particles is done is $PT$. Strictly speaking, the quantum limit is
only obtained for infinite large Trotter numbers $P$, however, for
practical purposes, it is usually sufficient to choose $PT$ in the order
of two times the Debye temperature.

In the following,
it is convenient to represent the position of the particles in reduced
dimensionless variables $\vec{r}_{nt}$ and a (symmetric)
matrix $h_{\alpha\beta}$
that contains the shape and the size of the simulation cell such that
\begin{equation}
R_{n\,t\,\alpha} = \sum_{\beta=1}^3 h_{\alpha\beta} r_{nt\beta}
\label{eq:transfo}
\end{equation}
with $0 \le r_{nt\alpha} < 1$ 
and $V = \det {\bf h}$. Spatial periodic boundary conditions are applied
by subtracting or adding unity to $r_{nt\alpha}$ once a molecular dynamics
step has moved $r_{nt\alpha}$ out of the allowed range. 
With this transformation,  the integration over
$\int d^3R_{11}\cdots\int d^3R_{NP}$  in 
Eq.~(\ref{eq:part_func}) can be replaced with  the expression
$V^{NP} \int d^3r_{11}\cdots\int d^3r_{NP}$.
This makes it possible to take the derivative of
$\ln Z(N,V,T)$ with respect
to the strain $\epsilon_{i}$. If we do not use the Voigt notation,
a (virtual) variation in the stress tensor 
$\delta \epsilon_{\alpha\beta}$ can be expressed as
\begin{equation}
\delta \epsilon_{\alpha\beta} = {1\over 2} \sum_{\gamma=1}^3 \left\{
 \left( {\bf h}^{-1} \right)_{\alpha\gamma} \delta h_{\gamma\beta} +
 \left( {\bf h}^{-1} \right)_{\gamma\beta} \delta h_{\alpha\gamma}
\right\}.
\label{eq:virtual}
\end{equation}
Combining Eq.~(\ref{eq:ela_def}) with Eqs.~(\ref{eq:part_func}) through
(\ref{eq:virtual}) then results in contributions to the elastic 
constants that resemble those obtained for classical systems.
In particular, $V_{\rm pot}$ in Eqs.~(\ref{eq:born}) and
(\ref{eq:correct}) has to be replaced with $V_{\rm eff}/P$, and the
factor $Nk_BT$ in Eqs.~\ref{eq:ideal} and \ref{eq:ideal_tensor}
has to be replaced with $Nk_BTP$.

If elastic constants are evaluated in an $NpT$ ensemble use can
be made of the relation
\begin{equation}
\left\langle \delta \epsilon_{\alpha\beta} 
             \delta \epsilon_{\beta\gamma} \right\rangle 
= (k_B T / V) \left({\bf C}^{-1}\right)_{\alpha\beta,\gamma\delta}
\label{eq:parrinello}
\end{equation}
where {\bf C} has to be represented as a $6 \times 6$ matrix, where we again
return to the Voigt notation.

\subsection{Application to Ideal Gas}
In order to discuss the expected statistical errors, it is helpful
to consider different contributions to the elastic constants. In order to do
this, we split the effective potential $V_{\rm eff}$ into two parts,
the real potential $V_{\rm pot}$ and the remaining part of the right
hand side of Eq.~(\ref{eq:pot_eff}), which we call $V_{\rm q}$.
The terms contributing to $C_{ij}$ can then be decomposed into the
Born contribution $C_{ij}^{\rm (Born)}$, a term 
$C_{ij}^{\rm (Born-q)}$, which is obtained by replacing
$V_{\rm pot}$ in the definition of $C_{ij}^{\rm (Born)}$
with $V_{\rm q}$, the term associated with the fluctuations of 
the generalized stress $C_{ij}^{\rm (fluc)}$, 
a similarly defined term $C_{ij}^{\rm (fluc-q)}$ stemming from
fluctuations of $\partial V_q /\partial \epsilon_i$, the cross
correlation $C_{ij}^{\rm (cross)}$ between the two terms 
$\partial V_{\rm q}/\partial \epsilon_i$ and 
$\partial V_{\rm pot}/\partial \epsilon_j$,
and the kinetic contribution $C_{ij}^{\rm (kin)}$.
\begin{eqnarray}
C_{ij} & =  &
C_{ij}^{\rm (Born)} + C_{ij}^{\rm (Born-q)} +
C_{ij}^{\rm (fluc)} + C_{ij}^{\rm (fluc-q)} 
\nonumber \\ & & +
C_{ij}^{\rm (cross)}  + C_{ij}^{\rm (kin)}
\label{eq:sum}
\end{eqnarray}
All terms related to $V_{\rm q}$ increase linearly in $P$ in leading order
as one can show in a particularly compact form if we exploit the harmonic
character of the springs in the ``ring polymers'',
by introducing appropriate normal coordinates
\begin{equation}
\tilde{R}_{nq\alpha} = {1\over\sqrt{P}} 
\sum_{t=1}^P {R}_{nt\alpha} e^{2\pi {\bf i} tq/P},
\end{equation}
so that $V_{\rm q}$ is diagonal in $\tilde{R}_{nq\alpha}$
\begin{equation}
V_{\rm q} = {1\over 2} \sum_{n=1}^N \sum_{\alpha=1}^3 
\sum_{q=1}^{P-1} k_q  \tilde{R}_{nq\alpha} \tilde{R}_{nq\alpha}^*
\end{equation}
with $k_q = 4m \sin^2(\pi q/P)\,/\,(\beta\hbar/P)^2$.

For the ideal gas we obtain:
\begin{eqnarray}
V C_{\alpha\beta\gamma\delta}^{\rm (kin)} & =  &
{1\over 2} 
{Nk_BTP} \delta_{\alpha\delta}  \delta_{\beta\gamma} \label{eq:ckin} \\
V C_{\alpha\beta\gamma\delta}^{\rm (Born-q)} & = &
{Nk_BT(P-1) } \delta_{\alpha\gamma}  \delta_{\beta\delta} \\
V C_{\alpha\beta\gamma\delta}^{\rm (fluc-q)} & = & -2
{Nk_BT(P-1) } \delta_{\alpha\gamma}  \delta_{\beta\delta},
\end{eqnarray}
while the estimator in the presence of a non-vanishing potential have the form
\begin{eqnarray}
V C_{\alpha\beta\gamma\delta}^{\rm (kin)} & =  &
{1\over 2 P} \delta_{\alpha\delta}  \sum_{nt} m 
\dot{R}_{nt\beta} \dot{R}_{nt\gamma} 
\label{eq:c_kin} \\
V C_{\alpha\beta\gamma\delta}^{\rm (Born-q)} & = &
{1\over 2 P}  \delta_{\alpha\gamma}  \sum_{it} \, k  \,
\delta R_{nt\beta} \, \delta R_{nt\delta}
\label{eq:c_born_q} \\
V C_{\alpha\beta\gamma\delta}^{\rm (fluc-q)} & = & -2
{Nk_BT(P-1) } \delta_{\alpha\gamma}  \delta_{\beta\delta},
\end{eqnarray}
with $\delta R_{nt\alpha} = R_{nt\alpha} - R_{nt+1\alpha}$.

In the presence of a non-vanishing $V_{\rm pot}$, the individual terms
will only be effected slightly, because in the quantum limit, changes
in the interaction potential are much smoother than the energy changes
due to changes in $V_{\rm q}$.
The net and properly symmetrized $C_{ij}$, however,  do
have well-defined values in the limit $P \to \infty$, which are, of course,
sensitive to the interaction potential $V_{\rm pot}$.
For the ideal gas, the symmetrized results are listed in
Table~\ref{tab:ideal}. In a numerical calculation, it clearly will 
be a problem that individual terms in Eqs.~(\ref{eq:ckin}-\ref{eq:c_born_q})
are of order $P >> 1$ while the final result is of order unity.

\subsection{Improved Primitive Estimators in PIMD}

Here, we want to suggest how one can easily improve on the statistical
properties of so-called primitive estimators in PIMD simulations.
As a first step we
define a function $\delta K_{\alpha\beta}$ which measures how far the averaged
tensor of the kinetic energy deviates from its expectation value
\begin{equation}
\delta K_{\alpha\beta} = 
 {1\over 2} \sum_{n,t} \left\langle \tilde{m}_{nt} v_{n\alpha} v_{n\beta} 
\right\rangle_{\rm simul}
- {1\over 2} Nk_B TP^2 \delta_{\alpha\beta},
\end{equation}
where $\tilde{m}_{it}$ represents the (kinetic) mass of particle $n$
at Trotter time $t$
(note that kinetic masses can be chosen arbitrarily
independent of the physical mass~\cite{tuckerman93})
and $v_{i\alpha}$ is the $\alpha$ component of its velocity.
$\langle \bullet \rangle_{\rm simul.}$  symbolizes an expectation
value obtained in a simulation.
Ideally, $\delta K_{\alpha\beta}$ would be zero, but, due to finite statistics,
finite time-steps, and other round-off errors, we will usually find 
$\delta K_{\alpha\beta} \ne 0$. It is very likely that this deviation 
$\delta K_{\alpha\beta}$ will convert into a similarly large
deviation in potential energy, in particular into $V_{\rm q}$
for large Trotter numbers $P$. At large $P$, the external potential is locally
only a small perturbation to the springs connecting neigbored beads on the
ring. Due to the conversion of kinetic energy to potential energy,
similar deviations from the exact
thermodynamic average can be expected
in $V_{\rm q}$. This makes it possible to define an optimized estimator
for quantities associated with $V_{\rm q}$, namely
\begin{eqnarray}
{1\over 2} \sum_n \sum_t {mP^2\over \beta\hbar} 
\left\langle\delta R_{nt\alpha} \delta R_{nt\beta} \right\rangle_{\rm optim.}
& = & \nonumber\\
{1\over 2} \sum_n \sum_t {mP^2\over \beta\hbar} 
\left\langle\delta R_{nt\alpha} \delta R_{nt\beta} \right\rangle_{\rm simul.}
& - & \delta K_{\alpha\beta}
\end{eqnarray}

Similar correction terms can easily be generalized to higher orders,
e.g. deviatians of 
$\langle \tilde{m}_{it}^2 v_{i\alpha}v_{i\beta}
 v_{i\gamma}v_{i\delta}\rangle_{\rm simu.}$
from its thermal expectation value 
can be expected to convert into 
$\langle (mP^2/\beta\hbar)^2  \delta R_{i\alpha}\delta R_{i\beta} 
\delta R_{i\gamma} \delta R_{i\delta}\rangle_{\rm simu.}$

\begin{figure}[hbtp]
\begin{center}
\leavevmode
\hbox{ \epsfxsize=80mm \epsfbox{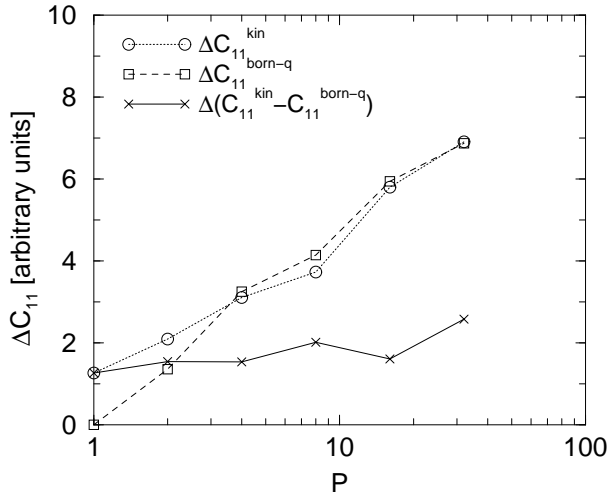} }
 \begin{minipage}{8.5cm}
  \caption{Statistical variances of the components 
   $C^{\rm (kin)}$ and $C^{\rm (Born-q)}$
  for varios Trotter numbers $P$. Similar conditions like same number of
  MD steps, same temperature, etc. have been used for all calculations.
  \label{fig:new_prim_est}
  }
 \end{minipage}
\end{center}
\end{figure}

In passing we want to explicitly give the improved
estimator for the kinetic energy,
which can be used in PIMD. It is similar to the so-called primitive
estimator~\cite{herman82}, but the expression $3Nk_BTP/2$ arising in
the original estimator is replaced by the actual average net kinetic energy.
With the quantities introduced in Eq.~(\ref{eq:sum}),
the average kinetic energy may simply be expressed as
\begin{equation}
  \langle T_{\rm kin} \rangle = V \sum_{\alpha=1}^3
   \left( C_{\alpha\alpha\alpha\alpha}^{\rm (kin)}
         -C_{\alpha\alpha\alpha\alpha}^{\rm (born-q)} \right)
  \label{eqn:new_est}
\end{equation}
As outlined above,
the new primitive estimator benefits from the correlation of kinetic and
potential energy in the isomorphic classical representation.
This is demonstrated in Fig.~\ref{fig:new_prim_est}, where the
average statistical deviation from terms of the type $C_{1111}$ are shown.

Note that the statistical error of 
$\Delta C_{11}^{\rm (Born-q)}$, which is identical to the statistical
error of the original primitve estimator, increases much slower than
the statistical error of the primitve estimator when examined
originally~\cite{herman82}.
This is due to the development of 
efficient sampling methods in the meantime~\cite{tuckerman93}
which suppress slowing down with increasing $P$.
The new estimator has a strongly reduced statistical noise as compared
to the original one, yet, results in the correct expectation value.
Furhtermore, using larger time steps than in our production runs, we
have noticed that systematic errors due to finite-time steps, are
considerably reduced as well with the new estimator. 
We want to note that in later production runs for strongly quantum
mechanical systems such as solid $^3$He, statistical error bars obtained
with the new estimators are as small as statistical error bars
asscociated with the so-called virial estimator~\cite{herman82}.

\subsection{Elastic Constants Under Pressure}
\label{sec:ela_press}

Care has to be taken when discussing elastic constants of systems under
pressure, because their definition is not unique.~\cite{wallace71}
One commonly distinguishes between the so-called Birch coefficients 
$B_{ij}$ and the elastic constants $C_{ij}$.
They are defined as the second derivative of the free energy with respect
to the Lagrangian strain and the ``regular'' strain, respectively.
The relation between $B_{ij}$ and $C_{ij}$ merely
depend on the externally applied stress but not on the symmetry of the
crystal. In the case of hydrostatic pressure,
the relations are given by~\cite{wallace71}
\begin{equation}
B_{ij} = C_{ij} + \Delta_{ij}
\label{eq:birch}
\end{equation}
with $\Delta_{ii} = -p$ for $1 \le i \le 6$,
$\Delta_{ij} = +p$ for $i \le 3$ and $j \le 3$ with 
$i \ne j$,
and zero else.

The generalization of our previos estimator of $C_{ij}$ to the 
non-zero pressure case are as follows: Evaluation of the strain fluctuations
as indicated in Eq.~\ref{eq:parrinello} result in the
Birch coefficients, while the formula given for zero-pressure $C_{ij}$ in the
$NVT$ ensemble enable us to calculate the $C_{ij}$ at non-zero pressures.

Note that the proper stability criterion for solids under pressure is a
positive definite matrix of Birch coefficients rather than a positive
definite matrix of elastic constants.~\cite{wang93} It is important to
keep in mind that symmetry relations of $C_{ij}$ which are 
obtained under the assumption of short-ranged two-body potentials
are not valid for
$B_{ij}$ in the case of any non-zero externally applied 
stress.~\cite{wallace71}

\subsection{Simulation Method}
\label{sec:method}
Simulations are carried out in both, the $NpT$ and the $NVT$ ensemble.
Molecular dynamics
are favorable over Monte Carlo simulations in the constant stress
ensemble $(NpT)$, because all variables are averaged simultaneously -
in particular all elements of the strain tensor. In order to keep
the externally applied stress tensor constant, the Parrinello-Rahman
method~\cite{parrinello81} has been applied to the classical representation
which is isomorphic to the quantum system. The classical representation
is defined by Eqs.~(\ref{eq:part_func}) and (\ref{eq:pot_eff}).
More details are given elsewhere~\cite{muser00}
on how to efficiently collapse the
time-scales associated with the different motions of the system
such as the intra-molecular breathing of the closed classical ring representing
the quantum point particle, the center-of-mass
motion of the ring, and the flucuations of the cell geometry.

\section{Applications}
\label{sec:applications}

\subsection{Argon}
Argon is a convenient test case for the calculation of quantum-mechanical
elastic constants, because it lies in between what is considered a
quantum solid like solid helium under pressure, and what is considered
a ``classical'' solid like solid Xenon. For the Argon test-case, we
intend to determine the shift of the quantum mechanical elastic
constants with respect to the classical elastic constants and the
relative
importance of the individual contributions to the net result.
The results will be an indicator for what we can expect from the
quantum mechanical shift in the elastic constants of other
solids.

\begin{figure}[hbtp]
\begin{center}
\leavevmode
\hbox{ \epsfxsize=80mm \epsfbox{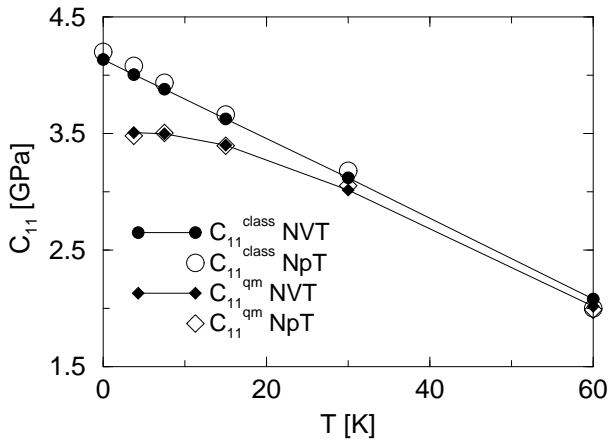} }
 \begin{minipage}{8.5cm}
  \caption{Elastic constant $C_{11}$ of solid Argon at ambient pressure as
  a function of temperatre $T$. Classical and quantum simulations were
  carried out at constant volume $V$ and constant external stress.
  \label{fig:ar_c_11}
  }
 \end{minipage}
\end{center}
\end{figure}

First, we compare classical to quantum mechanical calculations in
Fig.~\ref{fig:ar_c_11}, where results in the $NpT$ and the $NVT$ ensemble
are shown. In the simulations of Argon, a Lennard Jones potential
$V = 4 \epsilon [(\sigma/r)^{12}-(\sigma/r)^6]$ was
used with parameters $\epsilon = 1.67\times 10^{-21}$~J and 
$\sigma = 3.405\,\AA$. All simulations were based on system sizes $N = 500$
and statistics of $5\times 10^5$ time steps. The Trotter number in the
quantum runs has been chosen such that $PT = 120$. Increasing $P$ any
further does not change the results within the statistical error bars.

In Fig.~\ref{fig:ar_c_11}, it can be seen that there is a considerable
discrepency between classical and quantum mechanical elastic constants. 
The quantum $C_{11}$ levels at about $T = 15$~K and one may extrapolate
$C_{11} \approx 3.5$~GPa for the quantum mechanical ground state, while
classically, $C_{11} \approx 4.2$~GPa at zero temperature. The relative
effects in $C_{12}$ and $C_{44}$ are similar. Note that this effect of
a ${\cal O}(20\%)$ reduction in $C_{11}$ is considerably larger than the 
increase in
the lattice constant of about 1\% or the decrease of
10\% of the heat of formation. 

The bulk of the reduction in $C_{11}$ does not stem from the
expressions introduced in Sec.~\ref{sec:deriv}. There are two
important contributions, namely the Born term and the term related
to the fluctuation of the Lennard Jones potential. Details are
given in Table~\ref{tab:c_11_ar}. The classical system at
$T = 0$~K has only one non-zero contribution, namely
$C_{11}^{\rm (Born)} = 4.2$~GPa. As expected, differences between classical
and quantum results become negligible as the temperature approaches
(or surpasses) the Debye temperature, which in the case of Argon is
$T_D = 93$~K.

\subsection{Helium}

While Argon shows significant differences between classical and
quantum mechanical elastic constants, the corrections due to the kinetic
energies are farily small. Helium, in particular $^3$He,
is much more ``quantum'' than Argon and therefore a
more challenging test case than Argon.
The solid forms of $^3$He are only stable under pressure.~\cite{dobbs94} 
Hence, we have to distinguish between elastic constants and Birch coefficients.
There are three different, stable $^3$He lattice structures: the bcc phase
is mostly stable in the interval $0 \le T \le  2$~K at pressures
$3\,{\rm MPa} \le p \le 10$~MPa, hcp is the stable low-temperature phase at
pressures $p > 10$~MPa, and there is a small temperature regime at pressures
 $p > 150$~MPa where a stable fcc phase is found in between the hcp and the
fluid phase.

The simulations for helium are all based on the Aziz potential~\cite{aziz87},
which is known to be fairly reliable up to moderate pressures. Exchange
effects are neglected. In the temperature regime accessible to our simulation
they are generally believed to be unimportant. The transition to
a long-range ferromagnetic and antiferromagnetic transitions in hcp and
bcc $^3$He, respectively, only take place at temperatures in the
mK regime.

\subsubsection{fcc $^3$He}

In this subsection, we will focus on one particular representative point in
the stable fcc phase, because it is computationally very demanding to calculate
elastic constants. In order to be able to work with relatively small Trotter
numbers, it is necessary to go to large temperatures and low pressures.
Yet, we want to avoid to be too close to a phase transition, e.g., the fcc
fluid phase transition. The combination of $T = 18$~K and $p = 200$~MPa
seems to be an appropriate choice: The quantum limit is basically reached
with Trotter number $P = 32$ and the transition to the fluid phase
takes place at sufficiently higher temperature, namely at $T = 22$~K.
Although the thermodynamic stability field of the hcp phase is located
at temperatures $T < 18$~K for the applied pressure, one may certainly
expect metastability of the fcc phase in the time window accesible to our
simulations.

\begin{figure}[hbtp]
\begin{center}
\leavevmode
\hbox{ \epsfxsize=80mm \epsfbox{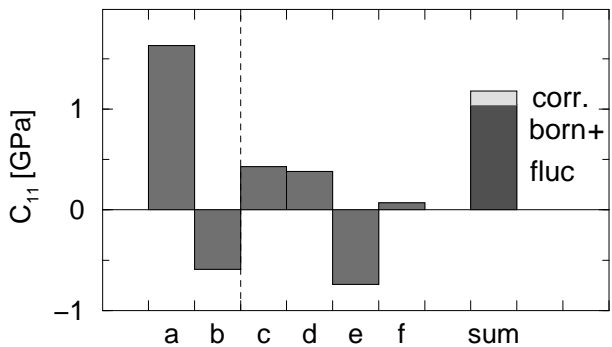} }
 \begin{minipage}{8.5cm}
  \caption{Individual contributions to $C_{11}$ of fcc $^3$He at pressure
  $p = 200$~MPa and temperature $T = 18$~K. 
  a: Born, b: fluc, c: kin,
  d: Born-q, e: fluc-q, f: cross; corr. summarizes c, d, e, and f.
  Trotter Number $P = 32$ and
  number of particles $N = 500$.
  \label{fig:c11_he_fcc}
  }
 \end{minipage}
\end{center}
\end{figure}

The different $C_{ij}$ along with their individual contributions
are listed in Table~\ref{tab:c_ij_fcc}. It is interesting to note
that the Cauchy relation $C_{12} = C_{44}$ for cubic crystals with
central potentials basically also hold for the quantum solid, e.g.,
$C_{44} / C_{12} \approx 0.95$, while the Birch coefficients don't:
$B_{44} / B_{12} \approx 0.50$. See Sec.~\ref{sec:ela_press} for the definition
and the relevance of Birch coefficients.
It is instructive to represent
the individual contributions to $C_{11}$ graphically, which is done
in Fig.~\ref{fig:c11_he_fcc}. It is noticable that the sum of the corrections
to $C_{11}$ which are related to the kinetic energy is relatively small,
while the individual contributions are fairly large. Yet, the solid
is far away from being classical. The kinetic energy of $^3$He is
about 88.9$~k_B$K, which is considerably larger than the thermal classical
energy of 1.5~k$_B \times 18$~K = 27~k$_B$K, thus 
$T/T_{\rm Debye} < 0.3$.
``Classical'' helium would
have $C_{11} = 1.63$~Gpa at the same external temperature and pressure.
From Fig.~\ref{fig:c11_he_fcc} we can see that it is possible to approximate
the elastic constants fairly reasonably if only the Born contribution
and the contribution due to the fluctuating real potential are included
into the calculation.

\subsubsection{bcc $^3$He}

\begin{figure}[thbp]
\begin{center}
\leavevmode
\hbox{ \epsfxsize=80mm \epsfbox{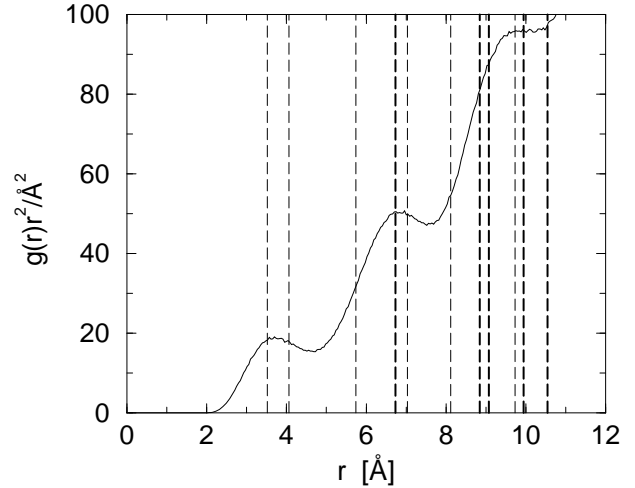} }
 \begin{minipage}{8.5cm}
  \caption{Radial distribution function $g(r)$ times $r^2$ as a function
  of distance $r$ for bcc $^3$He at $T = 2$~K and $P = 10$~MPa. The broken
  lines indicate the distance of nearest neighbors, next neighbors, etc.
  The width of the broken lines is proportional to the number of atoms in each
  shell.
  \label{fig:bcc_gr}
  }
 \end{minipage}
\end{center}
\end{figure}

The bcc phase can be considered to be the most interesting solid helium phase,
because it is classically unstable.
fcc and hcp phase can both be considered to be classically stable at low
temperatures, because their free energy differences are small.
An interesting
phenomena in the (quantum mechanical) bcc phase is the radial distribution
function $g(r)$:
The peaks in $g(r)$ can not easily be related to nearest neighbors, next
neighbors, etc., but the contributions of different neighboring shells 
``group'' together. This is shown in Fig.~\ref{fig:bcc_gr}. The final $g(r)$
can be interpreted as a sum of broadened individual lines, which are
represented in Fig.~\ref{fig:bcc_gr} as well. The overlap of such
broadened lines is accompanied by a strong diffusion of individual atoms. 
We did not investigate in depth this diffusion process.

In the case of $^3$He, some experimental data is available for
the elastic constants or Birch coefficients.
In Fig.~\ref{fig:cij_bcc}, we compare the $B_{ij}$ calculated in
our study with the available experimental data and
some theoretical predicitions.  Simulations at three different combinations
of pressures and temperatures were performed:
$p_1 = 10$~MPa with $T_1 = 2$~K, $p_2 = 7$~MPa with $T_2 = 1.5$~K, 
and $p_3 = 4$~MPa with $T_3 = 1$~K. In all cases, Trotter numbers
of $P = 256$ were found to reflect the quantum limit sufficiently well,
and the particle number was $N = 432$. The elastic constants can be assumed
to be mainly temperature independent, as the temperatures are far below
the Debye temperatures, e.g., the ratios $q_i = \langle T_{\rm kin} \rangle /
1.5 k_B T$ which are close to unity at the Debye temperature turned out
to be: $q_1 = 10.27$, $q_2 = 12.22$, and $q_3 = 15.4$.

For bcc $^3$He, the (relative) corrections in $C_{ij}$ which are related to 
the kinetic energy are much larger than in fcc $^3$He.  This can be seen in
Table~\ref{tab:c_ij_bcc_4}: the corrections make up nearly  80\%
of the total $C_{11}$ for a pressure $p = 4$~MPa and about 30\%
for a pressure of $p = 10$Pa (not explicitly shown in Tables).
We want to note that the elastic constants
obtained in the $NVT$ ensemble and in the $NpT$ ensemble agreed within
the statistical error bars.

\begin{figure}[hbtp]
\begin{center}
\leavevmode
\hbox{ \epsfxsize=80mm \epsfbox{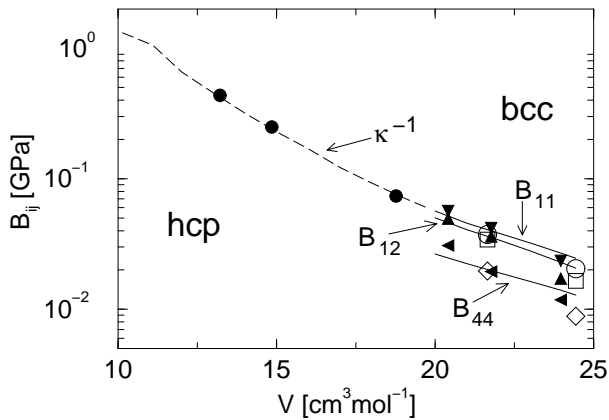} }
 \begin{minipage}{8.5cm}
  \caption{Birch coefficients for bcc $^3$He (right hand side)
   and bulk modulus for hcp $^3$He (left hand side)
  as a function of the molar volume. All filled symbols refer to this work.
  The broken line represents experimental data from 
  Ref.~[20].
  The solid lines represent theoretical predictions from 
  Ref.~[21].
  Open symbols refer to experimental data from
  Ref.~[22].
  \label{fig:cij_bcc}
  }
 \end{minipage}
\end{center}
\end{figure}

\subsubsection{hcp $^3$He}

hcp $^3$He has more independent elastic constants than bcc and fcc $^3$He,
respectively. The trend in all elastic constants, however,
is yet again the same as in bcc or fcc helium:
the larger the pressure the more dominant the ``classical'' contribution
to the elastic constants. Note that on the other hand,
the absolute corrections due to the kinetic energy
increase with increasing pressure. Details of the calculations are given
in Tables~\ref{tab:c_ij_hcp}-\ref{tab:c_ij_hcp_2}. 
Note that the classical elastic constants would be nearly three times
as large as the quantum mechanical elastic constants at a pressure
$p = 90$~MPa and even more than seven times as large in the case 
of $p = 15$~MPa.
Due to the good agreement with the experimentally
measured bulk modulus, see Fig.~\ref{fig:cij_bcc},
one may expect our quantum mechanical data to be fairly accurate.
To our knowledge, the
values presented in Tables~\ref{tab:c_ij_hcp}-\ref{tab:c_ij_hcp_2} 
are predictions. No theoretical
or experimental data is known to us.

\subsection{Kinetic energy of solid $^3$He}

The calculation and measurment of kinetic energies 
$\langle T_{\rm kin} \rangle$ in condensed
helium phases has attracted a lot of recent attention. Condensed
helium is highly quantum mechanical and therefore provides an ideal test
ground for the application of many-body quantum statistical theories.
Recent experimental developments
have made it possible to measure kinetic energies with great accuracy
and thus provide important tests on the theories.

\begin{figure}[hbtp]
\begin{center}
\leavevmode
\hbox{ \epsfxsize=80mm \epsfbox{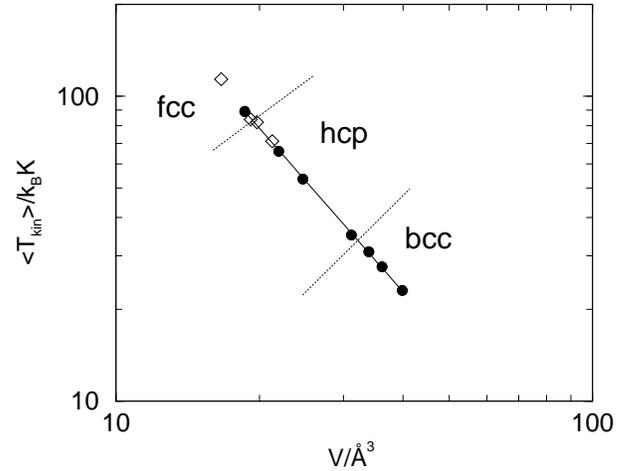} }
 \begin{minipage}{8.5cm}
  \caption{ Thermal kinetic energy $\langle T_{\rm kin} \rangle$
  as a function of the average volume per atom. Open diamonds are values from 
  PIMC simulations by Draeger and Ceperley. The straight line is a power
  law fit to our data. 
  \label{fig:t_kin_he}
  }
 \end{minipage}
\end{center}
\end{figure}

The data for $\langle T_{\rm kin} \rangle$ presented in this study
complement data by Draeger and Ceperley, which has been presented
recently~\cite{draeger00}. In particular, data for the bcc phase
and the low-pressure hcp regime are presented, see Fig.~\ref{fig:t_kin_he}.
We would like to emphasize that in both studies relatively high temperatures
were taken in the high-pressure regime with small
average atomic volume of $V < 20\,\AA^3$, e.g., the ratio of kinetic
energy to thermal energy, $q = 2 \langle T_{\rm kin} \rangle / 3 k_B T$
is only slightly larger than three. Of course, this is still well below
the Debye temperature, but some thermal activation will be present.

Our data can be very well fit with an expression of the type
\begin{equation}
\langle T_{\rm kin} \rangle = A V^{-\alpha},
\end{equation}
where $A$ and $\alpha$ are fitparameters. $A$ and $\alpha$ turn out to be
$\alpha \approx 1.78$ and $A = 16100$ if $V$ is expressed in
$\AA^3$ per atom and kinetic energies in units of k$_B$K.
Within the regime considered here, our data is reflected
within 1.5\% accuracy. In the bcc phase, this fit underestimated
$\langle T_{\rm kin} \rangle$ by about 1.5\%, in the hcp phase,
$\langle T_{\rm kin} \rangle$ is slightly overestimated.
The value of $\langle T_{\rm kin} \rangle$
near the quantum mechanical ground state seems to be mainly
a function of the molar or atomic volume only - relatively
insensitive to the actual crystalline phase.

We want to emphasize that the data points in Fig.~\ref{fig:t_kin_he}
which turn out to be larger than the values suggested by the fit,
all have relatively small values of the quantum parameter $q$.
Thus, going to even smaller values of $q$, which is computationally
expensive, might even increase the quality of the fit.
Omitting all data with $q < 5$, an exponent of $\alpha = 1.75$ is found.
It is quite surprising that the birch coefficients depend exponentially
on the molar volume (Fig.~\ref{fig:cij_bcc}),
the quantum mechanical ground state kinetic energy,
however, only changes algebraicly with $V$.

\section{Conclusions}
\label{sec:conclusions}

Expressions for the elastic constants of quantum solids have been presented
in terms of a path integral representation in the $NVT$-ensemble.
These expressions have then been applied and proven
useful in path integral molecular dynamics simulations of solid Argon and
hcp, fcc, and bcc helium III.  In the $NpT$ ensemble, the classical
formula can be used which relates strain fluctuations with elastic constants
or - in the case of non-zero external pressure - with Birch coefficients.
With the exception of hcp and bcc $^3$He, the $NVT$ expressions were
dominated by the terms which one knows from classical simulations: the
Born term and the potential fluctuation term. In the case of bcc $^3$He,
terms related to the kinetic energy dominate the elastic constants.

The quantum mechanical motion of the particles shows a stronger
effect on the elastic constants than one might expect. E.g., in the case of
Argon at zero external pressure, the quantum mechanical $C_{11}$ is reduced
by about 20\%, while cohesion energy and lattice constants are only
decreased by 10\% and increased by 1\%, respectively.
Hence, in order to have truely accurate estimates for elastic constants
from computer simulations,
quantum effects need to be taken into consideration.

While deriving the expressions for elastic constants, an improved
primitive estimator for the kinetic energy has been proposed. 
The statistical uncertainties of this estimator do not increase with
increasing Trotter number. Due to short correlation times,
the improved primitive estimator results in statistical error bars
smaller but in the order of the virial estimator for highly quantum systems
such as solid $^3$He.
However, unlike the virial estimator, the improved primitive
estimator can only be used in path integral
molecular dynamics, but not in Monte Carlo simulations.

\acknowledgments
We thank Kurt Binder for useful discussions.
Support from the BMBF through Grant 03N6015 and
from the Materialwissenschaftliche Forschungszentrum  
Rheinland-Pfalz is gratefully acknowledged.

\end{multicols}

\begin{table}
\begin{center}
 \begin{tabular}{c|ccc|c}
         & kin    & Born-q & fluc-q & sum\\ \hline
$C_{11}$ &  P     &   P-1  & -2(P-1)       &  1  \\
$C_{12}$ &  0     &    0   &   0     &   0 \\
$C_{44}$ &  P/2   & (P-1)/2&  -(P-1)     &  1/2   \\
\end{tabular}
\caption[]{
Individual components of $C_{ij}$ for the ideal gas in units
of $Nk_BT/V$ in the path integral representation.}
\label{tab:ideal}
\end{center}
\end{table}

\begin{table}
\begin{center}
\begin{tabular}{|c|c|c||c|c|c|c||c||c|}
 & Born & fluc & kin & Born-q & fluc-q & cross & corr & sum \\ \hline
quantum &3.72639&-0.30566&0.04337&0.03278&-0.05530&0.05469&0.07555&3.49627 \\
        &0.00005&0.00156&0.00001&0.00001&0.00053&0.00150&0.00137&0.00195\\ \hline 
class.  &4.03116&-0.15535&0.00278&0&0&0&0.00278&3.87860 \\
        &0.00004&0.00107&0.00001&&&&0.00001&0.00107 \\ 
\end{tabular}
\caption{Individual contributions to $C_{11}$ for Argon at $T = 7.5$~K
for the quantum and the classical calculation.
Lower rows give statistical incertainties based on 500.000 molecular dynamics
steps. The individual contributions are introduced in 
Eq.~(11).
The colon named ``corr'' summarizes all corrections involving contributions
from the kinetic energy: kin, Born-q, fluc-q, and cross.
}
\label{tab:c_11_ar}
\end{center}
\end{table}

\begin{table}
\begin{center}
\begin{tabular}{|c|c|c||c|c|c|c||c||c|}
 & Born & fluc & kin & Born-q & fluc-q & cross & corr & sum \\ \hline
$C_{11}$ & 1.63 & -0.59 & 0.43 & 0.38 & -0.74 & 0.07 & 0.14 & 1.18 \\
$C_{12}$ & 0.77 & -0.24 & 0 & 0 & 0.01 & 0.12 & 0.13 & 0.66 \\
$C_{44}$ & 0.77 & -0.23 & 0.21 & 0.19 & -0.36 & 0.05 & 0.09 & 0.63 \\
\end{tabular}
\caption{Individual contributions to $C_{ij}$ for fcc $^3$He at $T = 18$~K
and pressure $p = 200$~MPa. The colon named ``corr'' summarizes same terms
as in previos table.
Statistical error bars of final $C_{ij}$ are smaller than 5\%.
}
\label{tab:c_ij_fcc}
\end{center}
\end{table}

\begin{table}
\begin{center}
\begin{tabular}{|c|c|c||c|c|c|c||c||c|}
 & Born & fluc & kin & Born-q & fluc-q & cross & corr & sum \\ \hline
$C_{11}$ & 87.00 & -79.15 & 88.59 & 83.27 & -152.96 & 8.26 & 27.16 & 35.00 \\
\hline
$C_{12}$ & 35.19 & -29.47 & 0 &0& 4.84 & 16.44 & 21.28 & 27.00 \\
\hline
$C_{44}$ & 35.19 & -27.84 & 44.28 & 41.62 & -78.53 & 0.61 & 7.98 & 15.33 \\
\end{tabular}
\caption{Individual contributions to $C_{ij}$ for bcc $^3$He at 
$T = 2$~K and pressure $p = 4$~MPa.
The colon named ``corr'' summarizes same terms as in previos table.
Statistical error bars of final $C_{ij}$ are smaller than 10\%.
}
\label{tab:c_ij_bcc_4}
\end{center}
\end{table}

\begin{table}
\begin{center}
\begin{tabular}{|c|r|c|r|r|r|r|r|r|}
$p$ & $T$/K   & $C_{11}$  & $C_{12}$ & $C_{13}$ & $C_{33}$ & $C_{44}$ 
    & $C_{66}$  & $\langle V \rangle$/cm$^3$ \\ \hline
0.09 & 10.0   & 0.701     & 0.303    &   0.193 &  0.822    & 0.206
   &  0.193   &  13.22  \\
0.05 &  5.0   & 0.405     & 0.172    & 0.111   &  0.492    & 0.123
   & 0.108    &  14.85  \\
0.015 & 2.5    &  0.118    & 0.052    &  0.034  &  0.144    & 0.039
  &  0.032    &  18.77  \\
\end{tabular}
\caption{Independent elastic constants of hcp-$^3$He
as obtained in the $NpT$ ensemble.  The thermal expectation
value of the molar volume $\langle V \rangle$
is inserted as well. Pressure and elastic constants are expressed in
GPa. 
Statistical error bars of $C_{ij}$ are smaller than 5\%.
}
\label{tab:c_ij_hcp}
\end{center}
\end{table}

\begin{table}
\begin{center}
\begin{tabular}{|c|r|c|r|r|r|r|r|r|}
$V/$cm$^3$  & $T$/K   & $C_{11}$  & $C_{12}$ & $C_{13}$ & $C_{33}$ & $C_{44}$ 
    & $C_{66}$  & $\langle p \rangle$ \\ \hline
13.22 &  10.0 & 0.700     &  0.234   & 0.208    & 0.810    &  0.169
    & 0.192   & 0.090  \\
14.85 &   5.0 & 0.407     &  0.183   & 0.125    & 0.447    &  0.138
    & 0.120   & 0.050 \\
18.77 &   2.5 & 0.135     &  0.043   & 0.039    & 0.156    &  0.036
    & 0.025   & 0.015 \\
\end{tabular}
\caption{Independent elastic constants of hcp-$^3$He
as obtained in the $NVT$ ensemble.
The thermal expectation
value for the pressure $\langle p \rangle$
is inserted as well. $p$ and $C_{ij}$ are expressed in
GPa.
}
\label{tab:c_ij_hcp_2}
\end{center}
\end{table}

\begin{thebibliography}{99}
\bibitem{comment}
We use the symbol $C_{ij}$ for elastic constants in the Voigt notation
and $C_{\alpha\beta\gamma\delta}$ for elastic constants in the
tensor notation.
\end{thebibliography}
\end{document}